%% file: main.tex
\documentclass{INTERSPEECH2023}

\interspeechcameraready 

\usepackage{cleveref}
\usepackage{todonotes}
\usepackage{graphicx, subcaption}
\PassOptionsToPackage{hyphens}{url}\usepackage{hyperref}
\usepackage{hyperref}
\hypersetup{
  colorlinks=true,
  linkcolor=blue,
  urlcolor=cyan,
}
\title{Complex-valued neural networks for voice anti-spoofing}
\name{Nicolas M. M\"uller, Philip Sperl, Konstantin B\"ottinger}
\address{Fraunhofer AISEC}
  \email{nicolas.mueller@aisec.fraunhofer.de}

\begin{document}

\maketitle
 
\begin{abstract}
\input{tex/abstract}
\end{abstract}
\noindent\textbf{Index Terms}: voice anti-spoofing, audio deepfake detection, complex neural network 

\input{tex/content}

\newpage
\bibliographystyle{IEEEtran}
\bibliography{mybib}

\end{document}

%% file: tex/abstract.tex
Current anti-spoofing and audio deepfake detection systems use either magnitude spectrogram-based features (such as CQT or Melspectrograms) or raw audio processed through convolution or sinc-layers. Both methods have drawbacks: magnitude spectrograms discard phase information, which affects audio naturalness, and raw-feature-based models cannot use traditional explainable AI methods.
This paper proposes a new approach that combines the benefits of both methods by using complex-valued neural networks to process the complex-valued, CQT frequency-domain representation of the input audio. This method retains phase information and allows for explainable AI methods.
Results show that this approach outperforms previous methods on the "In-the-Wild" anti-spoofing dataset and enables interpretation of the results through explainable AI.
Ablation studies confirm that the model has learned to use phase information to detect voice spoofing.

%% file: tex/content.tex
\section{Introduction}
Artificial Intelligence (AI) is advancing at a rapid pace, and has enabled the development of numerous good and helpful applications that are changing the way we live, work, and interact with each other. However, along with these positive advancements, new threats have emerged, particularly in the form of deepfakes and spoofs.
Deepfakes are computer-generated images, videos, or audio that are created using AI algorithms to manipulate and alter original content. Although deepfakes can be used for harmless fun, they can also be misused for malicious purposes such as creating fake news, information campaigns, slander, or fraud. This is concerning as deepfakes can be highly convincing, making it challenging to distinguish between genuine and fake content.
Spoofing, on the other hand, involves creating a fake or manipulated version of oneself to circumvent biometric identification systems such as facial recognition or fingerprint scanners. This has significant implications, especially in the realm of security, as it can compromise the integrity of identity verification systems.
As a result, the detection of spoofed or faked content is becoming increasingly important. It is crucial to develop effective methods for detecting such misuse of AI in order to ensure that these technologies are used ethically and responsibly.

Using Machine Learning (ML) for voice anti-spoofing requires adequate preprocessing because 
human speech has long-range temporal dependencies, with one second of audio usually represented by $16,000$ single data points or more.
Previous approaches to voice anti-spoofing have taken one of two approaches to audio preprocessing. Some have used spectral preprocessing, transforming the time-domain waveform into a magnitude spectrogram using the short-time Fourier transform (STFT) or similar techniques. This approach has shown good performance but has one key disadvantage: it discards phase information from the spectrogram. This phase information is crucial for audio quality and voice naturalness~\cite{griffin1984signal}, and in the domain of speech-to-text (STT), it is laboriously recreated using Griffin-Lim or neural vocoders~\cite{griffin1984signal, oord2016wavenet, chen2020wavegrad}. As bad phase is very audible\footnote{Audio examples available at \href{https://google.github.io/phase-prediction}{google.github.io/phase-prediction}.}, we argue that it is a useful feature that should not be discarded.

The second approach to voice anti-spoofing is to design models that process raw audio directly,
which have been shown to be more effective~\cite{tak2021end, tak2021EndtoEnd, ge2021raw}, but lack transparency because explainable AI methods (XAI) such as saliency maps require input data with spatial dimensions (i.e. at least two-dimensional input).

In this paper, we propose a new approach to voice anti-spoofing that combines the benefits of both previous approaches. We transform input audio into a complex-valued spectrogram using the constant-Q transform, a technique closely related to the STFT, which yields a complex-valued frequency representation.
This representation is mathematically equivalent to the time-domain representation of the input audio, but more suited to machine-learning algorithms.
We then process this frequency representation using a complex-valued convolutional neural network, which allows us to process all information present in the audio file, without the need to discard the phase.
We demonstrate that our proposed approach outperforms both magnitude spectrogram-based models and raw-feature models. Our approach is conceptually and architecturally simple, which sets it apart from some of the recent state-of-the-art raw models. Moreover, it enables the use of explainable AI techniques.

In summary, our contributions to the field of voice anti-spoofing can be outlined as follows:
\begin{itemize}
\item We emphasize the significance of phase information in the complex Short-Time Fourier Transform (STFT) output, which has been neglected in previous research.
\item Based on this insight, we propose a corresponding complex-valued input feature and neural architecture for jointly processing both magnitude and phase information.
\item We evaluate our approach and demonstrate its superiority over both magnitude spectrogram-based models and raw-based models.
\end{itemize}
We 
provide an online web interface\footnote{\href{https://deepfake-total.com}{https://deepfake-total.com}} where one can test arbitrary input, including YouTube videos, against our model.

\section{Previous Work}


As previously mentioned, voice anti-spoofing traditionally employs one of two feature preprocessing approaches. The first approach involves using magnitude-spectral features like the constant-Q transform (CQT)~\cite{brown1991calculation}, log or mel-scaled spectrograms~\cite{stevens1937scale}. These techniques use the complex-valued STFT of a time-domain audio signal as a foundation, extracting the magnitude and discarding phase information. The result is then squared, scaled, and possibly binned using the mel-scale~\cite{stevens1937scale} to mimic the human ear's perception. Previously suggested voice anti-spoofing architectures have then employed neural components such as convolutions, attention mechanisms, recurrence, or transformer blocks~\cite{wang2021Comparative,lavrentyeva_audio_2017,lavrentyeva_stc_2019,afchar2018MesoNet,vaswani2017attention} to process the resulting spectrogram.
These models have shown good performance, but no longer achieve state of the art.

The second approach involves directly processing the time-domain ``raw'' audio using either neural convolutions on the audio file itself~\cite{chintha2020Recurrent} or stacks of sinc-layers~\cite{ravanelli2018speaker} corresponding to rectangular band-pass filters. 
Max-pooling is often used~\cite{tak2021EndtoEnd} to avoid phase mismatch between the audio waveform and the sinc wavelets. The result is then processed using stacks of convolutions and gated recurrence (RawNet2) \cite{tak2021EndtoEnd}, more advanced architectures like spectro-temporal graph attention neural networks (RawGat-ST) \cite{tak2021end}, or differentiable architecture search~\cite{ge2021raw}.

Although these ``raw'' models have been shown to deliver outstanding performance \cite{tak2021end, tak2021EndtoEnd, ge2021raw}, they lack explainability. This is because existing XAI techniques such as saliency maps~\cite{simonyan2014very} or Smooth Grad~\cite{smilkov2017smoothgrad} are designed for spatially-dimensioned input, while ``raw'' audio is a one-dimensional vector. As a result, it becomes challenging to comprehend and trust the models' judgments.
Among other disadvantages, this impedes the detection of ``learning shortcuts'' \cite{geirhos2020shortcut}, i.e. highly predictive, but artificial correlations between input and target which do not reflect real-world causality. 
Detecting such shortcuts is particularly crucial in voice spoof detection, where they have already been identified \cite{muller2021Speech} and may result in a lack of generalizability to real-world scenarios \cite{muller2022does}. Thus, there is an urgent demand for explainable anti-spoofing models.

\section{Proposed Approach}

\subsection{Complex-valued CQT spectrograms}
In digital signal processing, an input audio signal $X = [X[0], X[1], \ldots, X[n]]$ is represented in the time domain as a series of scalar values $X[t] \in [-1, 1] \in \mathbb{R}$, which correspond to measurements of air pressure at time $t/s$, where $s$ is the sampling rate (e.g., 16~kHz or 16,000 samples per second). To convert this signal into a frequency representation $Z$, one may use the Short Time Fourier Transform, which is a bijective transform:

\begin{equation}
    Z[t, k] = \sum_{n=0}^{N-1} W[n]X[n+t]e^{-2 \pi i k n N^{-1}}
\end{equation}
Here, $X$ is the time-domain signal, and $W$ is a window such as the Gaussian, Hann, or Hamming window of size $N$. 
The frequency-domain representation $Z$ is a complex-valued, two-dimensional vector, where $t$ is the temporal and $k$ the frequency index. 
Elements in $Z$ are complex-valued scalars, i.e. $Z[t, k] = \alpha e^{i \theta} \in \mathbb{C}$, where $\alpha \in \mathbb{R}$ is the magnitude and $\theta \in [0, 2\pi[$ the phase.

Although $Z$ is better suited for neural processing than the corresponding time-domain representation $X$, it also has its own drawbacks: the uniform time and frequency resolution for all frequency bins $k$ does not align with the non-linearities of the human auditory system~\cite{mceachern1992ear, moore2012introduction}.
To address this issue, we employ the constant Q transform (CQT)~\cite{brown1991calculation, blankertz2001constant} which processes the same number of cycles $Q$ for each frequency and spaces the frequency bins $k$ geometrically (i.e., higher frequencies are farther apart).
Therefore, for high frequencies $k$, CQT employs smaller window sizes $N_k$ (and thus higher time resolution), while for lower frequencies, it employs larger window sizes and thus larger frequency resolution.
Thus, instead of using the STFT, we convert time-domain audio $X$ into a complex-valued CQT spectrogram:
\begin{equation}
    Z_Q[t, k] = \frac{1}{N_k} \sum_{n=0}^{N_k-1} W_k[n] X[n+t] e^{-2 \pi i QnN_k^{-1}}
\end{equation}
Previous work has already shown that magnitude-based CQT is better suited to voice antispoofing than magnitude-based STFT~\cite{todisco2017constant}, which motivates our approach.

Finally, we convert the complex-valued CQT-spectrogram into a log representation by means of the following transformation:
\begin{align}
    \mathbb{C} & \to \mathbb{C}  \\
    |z| e^{i \theta} & \mapsto 
    \text{max} \left(\epsilon, -\text{log}_e |z| + c\right) \cdot \alpha \cdot e^{i \theta}
    \label{eq:logscale}
\end{align}
To account for the logarithmic perception of sound level by the human ear, we log-scale the magnitude of the complex number while keeping the phase intact.
However, $\text{log}_e |z|$ can be negative, because $\text{log}_e |z| e^{i\theta} = -\text{log}_e |z| e^{i\theta + i\pi}$, leading to ambiguity.
To address this, we apply a lower threshold of $\epsilon > 0$ to ensure that the magnitude remains positive.
Additionally, to ease model training, we introduce trainable parameters $\alpha, c \in \mathbb{R}^{> 0}$ to scale the magnitude.
In our experiments, these converge to approximately $\alpha=0.15$ and $c = -0.3$.
To summarize, we utilize CQT followed by \cref{eq:logscale} to generate complex-valued CQT log-spectrograms (C-CQT).


\subsection{Complex-valued neural networks}
Complex-valued neural networks (CVNN) are networks where both inputs $X$ and weights $\theta$ are complex numbers $a + ib \in \mathbb{C}$. Apart from that, CVNNs rely on many of the same fundamental components as traditional real-valued ones. For instance, linear layers, convolutions, and simple recurrent networks like RNN operate identically in both types of networks. However, more complex components such as attention and recurrence, such as GRU~\cite{cho2014properties}, need to be adapted, since typical activation functions like \emph{sigmoid}, \emph{tanh}, and \emph{softmax} cannot be applied directly to complex-valued scalars \cite{bassey2021survey}.
For example, the complex \emph{ReLU} function can be implemented as
\begin{equation}
    CReLU(x) = \text{max}(0, \Re(x)) + i \text{max}(0, \Im(x))
\end{equation}
where $\Re$ and $\Im$ denote the real and imaginary parts of a complex number.
In supervised classification, class scores $y$ for $n$ classes can be derived from complex-valued logits $z$ as follows:
\begin{equation}
    \text{softmax}(|z|)_i = \frac{e^{|z_i|}}{\sum_{j=1}^n e^{|z_j|}}
    \label{eq:act}
\end{equation}
Classes with high probability are represented by logits with a large magnitude. These models, like their real counterparts, are trained using gradient descent~\cite{bassey2021survey} through the use of Wirtinger Calculus~\cite{wirtinger1927formalen}.

\subsection{Proposed Architecture}
\begin{figure}
    \centering
    \includegraphics[width=0.22\textwidth]{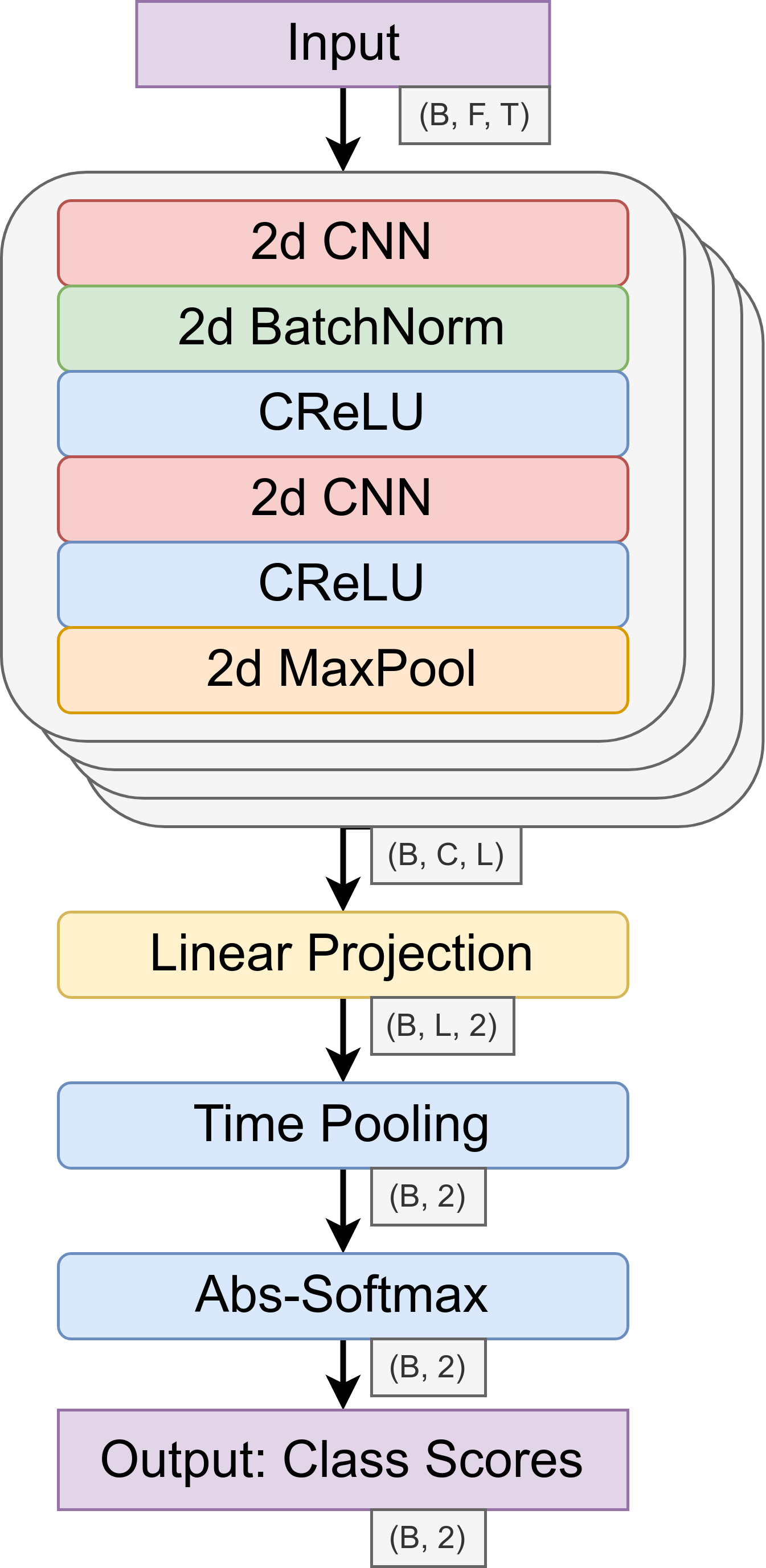}
    \caption{Our proposed complex-valued architecture. }
    \label{fig:architecture}
\end{figure}

In order to process complex-valued, frequency-domain audio input, we use a complex convolutional network.
It consists of four blocks of convolutions, where each block is composed of two-dimensional complex convolutions (kernel size $3$, stride $2$ and padding $1$), complex ReLU and complex Batch-Normalisation.
This is followed by linear projection, which consists of three complex-valued linear layers with CReLU activation and Dropout of $40\%$. This is followed by two-dimensional complex time pooling.
The output is a two-class complex vector of shape $(B, 2)$, to which we apply \cref{eq:act} in order to obtain real-valued class scores.
There are two target classes: class $0$, which corresponds to bonafide audio, and class $1$, which corresponds to spoofed audio.
The input frequency-domain audio has shape $B, F, T$ where $B$ is the batch-size, $F$ the number of frequency bins, and $T$ the number of time bins. The convolutional stack outputs a tensor $B, C, L$ where $C$ is the number of output channels and $L$ is the aggregated time.
\Cref{fig:architecture} provides an overview.

\section{Evaluation}
\subsection{Datasets}
We are utilizing two datasets for our experiments. The first dataset is the Logical Access (LA) section of the ASVspoof 2019 dataset~\cite{todisco2019asvspoof}. This dataset is widely used in related research~\cite{tak2021end,tak2021EndtoEnd,ge2021raw,wang2021Comparative,lavrentyeva_audio_2017,lavrentyeva_stc_2019,afchar2018MesoNet,vaswani2017attention} and is considered one of the most established anti-spoofing datasets.
It includes audio files that are either genuine human speech recordings (bona-fide) or manipulated/synthesized audio (spoofed). The spoofed audio files are generated using 19 different Text-To-Speech (TTS) synthesis algorithms that are considered a threat to the authenticity of human voice in terms of spoofing detection. Therefore, the spoofed audio files are labeled as ``attacks'' by ASVspoof. The dataset consists of 19 attackers, labeled as A1 to A19, and for each attacker, there are 4914 synthetic audio recordings and 7355 genuine samples.

Furthermore, we utilize the ``In-the-Wild'' (ITW) dataset~\cite{muller2022does}, which comprises $37.9$ hours of audio recordings, either spoofed ($17.2$ hours) or authentic ($20.7$ hours), obtained from $58$ English-speaking celebrities and politicians via online video sharing platforms, such as YouTube.
It tries to match the speaking style for each pair of spoofed and authentic instance to prevent the model from distinguishing spoofs based on extraneous characteristics (also called ``machine learning shortcuts''), such as speaking style, background noise, or setting.
For instance, for a spoofed speech by Barack Obama, an authentic speech was also included.
Our objective is for the model to distinguish spoofed audio based on either disfluencies that indicate poor naturalness or artifacts resulting from the text-to-speech (TTS) process.

Since our primary concern is the real-world performance of the models on unseen audio recordings, we train them on the entire ASVspoof dataset (all splits) and assess their generalization abilities using the "In-the-Wild" dataset.

\begin{table*}[h!]
    \centering
    \caption{Model evaluation results on the ``In-the-Wild'' dataset. The results are averaged over three independent runs, with the standard deviation displayed.}
    \input{res/results}
    \label{tab:res}
\end{table*}

\subsection{Data-augmentation and Hyperparameters}
Previous research has identified a significant learning shortcut in the ASVspoof 2019 dataset, whereby the duration of silence in an audio file is highly correlated with its corresponding label. This correlation has led to an overestimation of anti-spoofing system performance~\cite{muller2021Speech}. To obtain a realistic estimate of the anti-spoofing model performance, it is crucial to eliminate this shortcut. 
We address this issue by following the suggestions made by related work~\cite{muller2021Speech, zhang122021effect} and trim the beginning and end silences of an audio file. Moreover, we randomly select a 2-second segment from the remaining audio to ensure that all audio files have equal length. This prevents our models from using the presence or position of the cut-off as a new learning shortcut. We apply this technique to all subsequent analyses.

In addition, inspired by the widespread use of data augmentation in the image domain, we employ the following audio data augmentations in all subsequent analyses. By doing so, we further aim to avoid shortcuts, model overfitting, and promote model generalization.

First, we use Adversarial Retraining~\cite{goodfellow2014explaining}. Namely, we employ the Fast Gradient Sign Method (FGSM) to craft adversarial examples with a probability of $20\%$ and a magnitude of $0.25\%$ of the input magnitude and supply these as additional training examples. This method has been shown to improve model generalization~\cite{goodfellow2014explaining} and regularization~\cite{Regulari30:online}.
Second, we use audio augmentations~\cite{iver56au88:online} such as room impulse response, frequency masking, pitch- and timeshift, gain, high, low- and bandpass filters, polarity inversion, and clipping to diversify the input data with a probability of $20\%$.
Third, again with a probability of $20\%$, we partially encode the input using MP3, ulaw, and alaw codecs to perform compression.
Lastly, we add random clips of noise and music to the audio files with a probability of $5\%$ by utilizing the MusDB18~\cite{musdb18} and Noise ESC 50~\cite{piczak2015dataset} datasets.

We utilize the Adam optimizer with a learning rate of $5 \cdot 10^{-3}$ to train all models. The models are trained for $25$ epochs with a weight decay of $10^{-6}$ and a batch-size of $32$.
To prevent overfitting, we employ early stopping with a patience of $3$ epochs and a minimum delta of $5 \cdot 10^{-4}$. We implement the models using Python and PyTorch.
For the Constant-Q Transform (CQT), we use a 
hop size of $32$ samples with the ``Hann'' windowing function.
We train all models using an Nvidia DGX-A100 server, which is equipped with eight A100 GPUs, each having a memory of 40GB, and a total server memory of 1024GB. It features two AMD Rome 7742 processors, each with 64 cores.


\subsection{Evaluation and Results}


We adopt the Equal Error Rate (EER) as the performance metric for our model evaluation, consistent with previous studies~\cite{tak2021end,tak2021EndtoEnd,ge2021raw,todisco2019asvspoof,nautsch2021ASVspoof}. 
EER is the point on the Receiver Operating Characteristic (ROC) curve where the false acceptance rate and false rejection rate are equal. 
For each experimental configuration, we conduct three independent trials and report the mean and standard deviation of the EER. 
\Cref{tab:res} summarizes the results of our proposed model compared to models from related work.
We observe the following: Firstly, our complex-valued model demonstrates an absolute improvement over related work by about $3\%$ EER when compared to ``raw'' features and $10\%$ EER when using magnitude-based CQT features. Secondly, we observe the least amount of overfitting, as indicated by the small gap between the training and testing performance.

\subsection{Ablation Study}
In order to evaluate the impact of the phase information, we perform an ablation study where we keep all parameters for our proposed model, but discard or randomize the phase information from the complex spectrogram.
\Cref{tab:abl} presents the results.
We observe that the removal of phase information ("zero phase") results in a $4\%$ degradation in model performance with respect to EER. This suggests that the proposed model has indeed learned to detect spoofing attacks by utilizing the mismatch between phase and magnitude information to some extent.
Furthermore, randomizing the phase information causes an additional $4\%$ decrease in performance with respect to EER. We presume that this is due to the fact that the model not only lacks phase information, but also has to cope with a significant amount of noise when dealing with randomly generated phase information.

\begin{table}[]
    \centering
    \caption{Ablation study results for the C-CQT feature: full phase retention, random phase selection, or zero assignment.}
    \resizebox{0.48\textwidth}{!}{%
    \input{res/ablation}
    }
    \label{tab:abl}
\end{table}

\subsection{Explainable AI}

\begin{figure}
    \centering
        \includegraphics[width=0.45\textwidth]{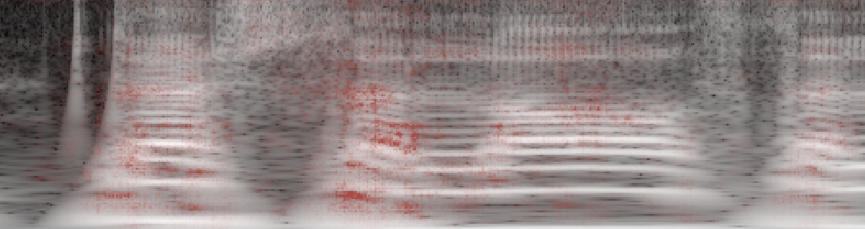}
    \caption{SmoothGrad~\cite{smilkov2017smoothgrad} applied on an instance from the ASVspoof dataset.
    }
    \label{fig:xai}
\end{figure}

Our model allows the application of explainable AI techniques such as saliency maps~\cite{simonyan2014very} and Smooth Grad~\cite{smilkov2017smoothgrad}. These techniques allow us to visualize how our model processes audio in the frequency domain, thereby enabling us to gain insight into the decision-making process of our model.
To demonstrate the effectiveness of these techniques, we apply Smooth Grad 
and present an exemplary output in \Cref{fig:xai}. Our analysis shows that the model does not rely on any obvious shortcuts, such as the duration of silence\cite{muller2021Speech} or other artifacts such as sampling or ringing issues in the upper frequency bands. Instead, the model focuses on the speech signal itself, as evidenced by the red XAI-overlay coinciding with the spectro-temporal bins containing descriptive features.

\section{Conclusion and Future Work}
This paper introduces a novel approach to voice anti-spoofing, which involves utilizing complex-valued features and complex-valued neural networks. This approach is motivated by the fact that magnitude-based feature extraction discards phase information, which is difficult to spoof yet crucial for producing natural-sounding synthesized speech, thereby constituting a useful input feature to voice anti-spoofing systems.

Building on previous research, we thus propose using the complex-valued Constant-Q Transform (C-CQT) as a new input feature and suggest a corresponding complex-valued neural architecture. 
Our proposed approach not only outperforms previous related work but also allows for explainability through established XAI-methods, which is lacking in ``raw''-feature-based models.
Through ablation studies, we demonstrate that our model has effectively learned to use phase information to distinguish between spoofed and bona-fide audio. These results indicate the potential of our approach in improving the accuracy of voice anti-spoofing and advancing the understanding of the underlying decision-making process.



In summary, we suggest an effective architecture that can be useful for anti-spoofing purposes. Despite the simplicity of our current model, it already surpasses the performance of related works that employ more sophisticated components. We are confident that incorporating further model refinements would be straightforward and could potentially enhance the ability to detect spoofed voice signals.

%% file: res/results.tex

\begin{tabular}{ll|ll|ll}
\toprule
                            Model Name & Feature Type   &  EER Test ITW         &EER Train & Epoch Time (s) & Num. Parameters              \\
\midrule
Proposed Model                          &        C-CQT &  \textbf{26.95±3.12}  & 12.382±6.58 & 276.6±6.5  &  2,459,492  \\
CRNN Spoof \cite{chintha2020Recurrent}  &          Raw &          29.90±4.33   &  8.103±0.59 &  177.0±5.0 & 3,330,562           \\
MesoNet \cite{afchar2018MesoNet}        &          CQT &     37.05±3.93        & 11.496±0.19 & 187.0±21.2 & 11,698              \\
RawGat-ST \cite{tak2021EndtoEnd}        &          Raw &   39.29±2.41          & 11.802±0.12 & 402.5±1.1  & 440,810             \\
RawNet2 \cite{tak2021end}               &          Raw &      40.00±0.25       & 11.680±0.02 &  113.4±5.9 & 17,648,770          \\
LCNN \cite{chintha2020Recurrent}        &     Mel-Spec &         55.07±7.09    & 21.575±1.82 & 170.8±12.2 & 178,306             \\
Deep ResNet~\cite{alzantot2019Deep}     &     Mel-Spec &  58.80±0.23           & 36.21±0.65  & 192.7±6.8 & 111,874 \\
\bottomrule
\end{tabular}

%% file: res/ablation.tex

\begin{tabular}{llll}
\toprule
    Model Name &  Phase & EER Test ITW &   EER Train \\
\midrule
Proposed Model &                    full &     \textbf{26.95±3.12} & 12.38±6.58 \\
 &                                  zero &              31.29±1.31 &  5.94±0.37 \\
 &                                  random &            35.79±0.13 &  8.09±0.18 \\
\bottomrule
\end{tabular}